# POSSIBILITIES OF SIMULATION OF FLUID FLOWS USING THE MODERN CFD SOFTWARE TOOLS

*Alexey N. Kochevsky*

*Research Scientist, Department of Applied Fluid Mechanics,*
*Sumy State University,*
*Rimsky-Korsakov str., 2, 40007, Sumy, Ukraine*
*alkochevsky@mail.ru*

**Abstract:** The article reviews fluid flow models implemented in the leading CFD software tools and designed for simulation of multi-component and multi-phase flows, compressible flows, flows with heat transfer, cavitation and other phenomena. The article shows that these software tools (CFX, Fluent, STAR-CD, etc.) allow for adequate simulation of complex physical effects of different nature, even for problems where performing of physical experiment is extremely difficult.

## INTRODUCTION

Fluid flows play the key roles in the working process of many modern engineering devices. Designing of these devices for the required operational parameters is impossible without reliable prediction of characteristics of these flows. As many modern engineering devices are expensive and labor consuming in manufacturing, their physical modeling with experimental determination of their working parameters at different modes, as a rule, requires large temporal and financial expenses. Besides, due to restricted possibilities of modern experimental sensors and measuring instruments, experimental observations do not give complete information about the investigated phenomena. Due to the nature of fluid medium itself, fluid flows often occur in very complex way, with presence of transitional effects, stagnation zones and vortex structures, and at supersonic velocities – also with generation of compression shock waves. The situation is still more complex when heat transfer is present, when considering flow of a mixture of several substances, free surface flows, flows with suspended particles, flows with cavitation, boiling, condensation, combustion, chemical reactions.

All these factors explain the growing interest for the software tools for simulation of fluid flows allowing for prediction of characteristics of these flows and working parameters of engineering devices at the stage of designing, before their manufacturing. The branch of science dealing with simulation of fluid flows with heat and mass transfer in various engineering and natural objects is computational fluid dynamics (CFD).

As the computational power of computers grew and, at the same time, their prices became affordable for more and more users, since the 70ies of XX century, rapid development of commercial CFD software has started. Till the beginning of 90ies, this software could be installed only at powerful computers like workstations. In 90ies, the cheap personal computers (PC) have become as powerful as those workstations, and CFD software packages designed for PC have appeared.

Nowadays, dozens of software packages intended for solution of fluid flow problems are available. A lot of them are listed, e.g., at the site www.cfd-online.com. Among the CFD software most recognized worldwide the following packages should be mentioned: CFX (www-waterloo.ansys.com/cfx/), Fluent (USA, www.fluent.com), STAR-CD (England, www.cd-adapco.com, www.adapco-online.com), Numeca (Belgium, www.numeca.be), etc. The packages FlowER (Ukraine, www.flower3d.org) and FlowVision (Russia, www.tesis.com.ru, www.flowvision.ru, see examples of computation also in [1]) are also worth to be mentioned.

Initially, computational fluid dynamics was developed for solution of problems of aerospace industry – simulation of processes in combustion chambers of rocket engines, simulation of physical-chemical processes in the flow around rocket airframe and supersonic aircrafts. Nowadays, field of application of CFD is essentially extended by civil engineering problems.

According to information presented at those sites, we list some most important fields of application and problems solved with methods of CFD using the commercial software. Transport (by land, by sea, by air): prediction of resistance caused by upstream air or water flow, ventilation and fire safety of passenger compartments, simulation of fuel combustion in combustion chambers. Fluid machinery: prediction of performance curves and operation modes of pumps, compressors and turbines at different configuration of flow passage, prediction of cavitation phenomena. Foundry: simulation of the founding process. Construction: prediction of wind loads on buildings, ventilation and fire safety of apartments. Power engineering: simulation of combustion flows in the burners and boilers of heat power plants. Ecology and natural disasters: simulation of distribution of pollution in water and air, simulation of expansion of fire in forests and towns.

Methods of CFD assume computation of liquid and gas flows by numerical solution of Navier – Stokes and continuity equations (for turbulent flows – Reynolds equations) which describe the most general case of movement of fluid medium. The correspondent sequence of actions, from creation of a geometrical model and specifying boundary conditions to analysis of computation results, is described in the paper [2]. The present paper describes the additional model equations to be included into the set of model equations for simulation of fluid flows of multi-component and multi-phase media, compressible flows, flows with heat transfer, cavitation, etc. These additional model equations are described as they are implemented in the leading software tools.

In general, tendency of development of the leading CFD software tools is implementation of a set of mathematical models in each of them, allowing for simulation of all the physical phenomena that may occur in practice as full as possible. A user turns on the necessary model equations while setting a problem, in several mouse clicks, and then specifies relevant boundary conditions and other required data. In this paper, the examples of problems are also listed that can be solved by inclusion of correspondent model equations.

SIMULATION OF INCOMPRESSIBLE FLUID FLOWS

*1. Laminar flow.* In the modern CFD software tools, the computation of liquid or gas flow is performed by numerical solution of system of equations that describe the most general case of movement of fluid medium. These equations are of Navier – Stokes (1) and continuity (2):

$$\frac{\partial}{\partial t}(\rho u_i) + \frac{\partial}{\partial x_j}(\rho u_i u_j) = -\frac{\partial p}{\partial x_i} + \frac{\partial}{\partial x_j}\left[\mu\left(\frac{\partial u_i}{\partial x_j} + \frac{\partial u_j}{\partial x_i}\right)\right] + f_i, \tag{1}$$

$$\frac{\partial \rho}{\partial t} + \frac{\partial}{\partial x_j}(\rho u_j) = 0. \tag{2}$$

For these equations, a brief form of record is used here. The summation on the same indices is assumed, $i, j = 1 \ldots 3$, $x_1, x_2, x_3$ – coordinate axes, $t$ – time. Full form of record for these equations in curvilinear coordinate system is presented, e.g., in [3]. The term $f_i$ expresses the action of body forces.

In this set of 4 equations, independent parameters being sought are 3 components of velocity $u_1, u_2, u_3$ and pressure $p$. Density $\rho$ of liquid as well as gas under velocities below 0.3 of Mach number is assumed to be constant.

The boundary conditions are posed usually as follows. Zero velocities are set at all the solid walls. At the inlet section, the distribution of all the velocity components is specified. At the outlet section, first derivatives of velocity components (in the direction of flow) are assumed to be zero. The pressure is present in the equations (1) only in first derivatives, thus, the user needs to specify pressure only at any arbitrarily selected node of the computational domain.

*2. Turbulent flow.* As a rule, flows in engineering are turbulent. Direct modeling of turbulent flows by numerical solving of Navier – Stokes equations, written for instant velocities, is still extremely difficult. Besides, as a rule, of interest are usually time averaged and not instant velocity values. Thus, for analysis of turbulent flows, instead of Navier – Stokes equations (1), Reynolds equations (3) are used:

$$\frac{\partial}{\partial t}(\rho \overline{u_i}) + \frac{\partial}{\partial x_j}(\rho \overline{u_i u_j}) + \frac{\partial}{\partial x_j}(\rho \overline{u'_i u'_j}) = -\frac{\partial p}{\partial x_i} + \frac{\partial}{\partial x_j}\left[\mu\left(\frac{\partial \overline{u_i}}{\partial x_j} + \frac{\partial \overline{u_j}}{\partial x_i}\right)\right] + f_i, \qquad (3)$$

where $\overline{u_1}$, $\overline{u_2}$, $\overline{u_3}$ – time averaged velocity components,

$\overline{u'_1}$, $\overline{u'_2}$, $\overline{u'_3}$ – fluctuating velocity components.

Different turbulence models are used for closure of these equations. These models are reviewed, e.g., in [2]. Besides, many of the physical processes considered below have essential influence on the turbulence, and for simulations of those flows the recommendations stated in the corresponding papers should be taken into account.

## SIMULATION OF SINGLE-PHASE MULTI-COMPONENT FLOWS

*1. Flow of mixture of several incompressible media.* When considering flow of mixture of two or several liquids (or gases – at small velocities of flow) with different densities, the density of mixture $\rho_m$ becomes variable and depends on concentration. In order to calculate the concentration of some component of mixture in each cell of space, the system of model equations is supplemented with one additional differential equation in partial derivatives (DEPD) – equation of transfer of concentration $C$:

$$\frac{\partial}{\partial t}(\rho C) + \frac{\partial}{\partial x_j}(\rho u_j C) = -\frac{\partial J_j}{\partial x_j} + R_r, \qquad (4)$$

where $R_r$ – rate of dilution process or chemical reaction, if available.

The term $J_j$ expresses intensity of molecular diffusion along the coordinate $x_j$ and is related to the concentration $C$ according to the second Fick's Law:

$$J_j = -D\frac{\partial C}{\partial x_j},$$

where $D$ – diffusion coefficient.

When $n$ fluid components with densities $\rho_1, \rho_2, …, \rho_n$ are mixed, in order to simulate the mixing process, the system of model equations should be supplemented with $n$ equations (4), one own equation for each component. Imagine, for example, a vessel filled with fluid 1, into which other fluids (2, 3, …, $n$) by several pipes are supplied, pumping the fluid 1 away. At the initial moment, the concentration of fluid 1 in the vessel is 1, the concentration of other components is 0. In the pipes supplying the fluids to the vessel, on the contrary, the concentration of corresponding fluid is 1. As a result of computation, after each time step, the values of concentration $C_1, C_2, …, C_n$ of each component in each cell of space will be obtained.

The density of mixture $\rho_m$ is computed as follows:

$$\frac{1}{\rho_m} = \frac{C_1}{\rho_1} + \frac{C_2}{\rho_2} + … \frac{C_n}{\rho_n}, \qquad (5)$$

where $C_1 + C_2 + … C_n = 1$.

The viscosity of mixture $\mu_m$ is computed as follows:

$$\mu_m = C_1 \mu_1 + C_2 \mu_2 + … + C_n \mu_n, \qquad (6)$$

where $\mu_1, \mu_2, …, \mu_n$ – viscosity of each component. When computing flows with heat transfer, specific heat conductivity and heat capacity of the mixture are computed the same way.

The equation of concentration transfer (4), written for time averaged values for simulation of turbulent flows, looks as follows:

$$\frac{\partial}{\partial t}(\rho \overline{C}) + \frac{\partial}{\partial x_j}(\rho \overline{u_j}\overline{C}) + \frac{\partial}{\partial x_j}(\rho \overline{u'_j C'}) = -\frac{\partial \overline{J_j}}{\partial x_j} + \overline{R_r}. \qquad (7)$$

*2. Flow of mixture of several compressible media.* If a gas mixture flows with velocities comparable with the sonic speed for the corresponding media, the system of model equations required for simulation of the process should include the energy equation. This approach is described with more details in the next chapter.

SIMULATION OF FLOWS WITH HEAT TRANSFER AND COMPRESSIBILITY

***1. Variation of density and temperature of stationary medium under forced compression.*** Density of liquid, except for some special cases (e.g., hydroblow), can be considered as independent on pressure. Density of gas can be changed by forced compression in a closed volume. The equation that describes the dependence of temperature and density on pressure is known as the equation of state:

$$p = \rho R T, \tag{8}$$

where $R = 287$ joule / (kg K) – universal gas constant.

The equation of state (8) is true for the perfect gas, i.e., when the interaction between molecules occurs by elastic collisions, and the linear size of molecule is small in comparison with average intermolecular distance. At very low temperature and/or high pressure, use of this equation can cause substantial discrepancy with the experimental observations for the real gas. In order to get more exact results in this case, e.g., the equation of state suggested by Van der Waals may be used:

$$\left(p + A\rho^2\right)\left(\frac{1}{\rho} - B\right) = RT, \tag{9}$$

where

$$A = \frac{27}{64}\frac{R^2 T_c^2}{p_c}, \qquad B = \frac{RT_c}{8 p_c},$$

$p_c$ and $T_c$ – pressure and temperature corresponding to the phase transition.

***2. Simulation of flows with heat transfer in case of incompressible or weakly compressible fluid.*** Temperature of fluid can change due to heat conductivity, when the fluid is in contact with a certain object (e.g., solid walls) which temperature differs from the temperature of fluid, or due to some processes inside fluid accompanied by heat generation.

In flows of incompressible and weakly compressible fluid (at flow velocities below 0.3 Mach number), density of fluid depends only on temperature, and influence of pressure differences on variations of density and temperature is negligible. From the computational point of view, in order to account for temperature variations, the system (1) – (2) should be supplemented with an additional DEPD – the energy equation:

$$\frac{\partial}{\partial t}(\rho H) + \frac{\partial}{\partial x_j}(\rho u_j H) = -\frac{\partial Q_j}{\partial x_j} + f_i u_i, \tag{10}$$

where the term $f_i$ expresses action of body forces.

In this equation, internal heat sources, related, e.g., with chemical reactions, are not included. Full enthalpy $H$ is coupled with full energy $E$, internal energy $e$ and static or specific enthalpy $h$ with the following relations:

$$H = E + \frac{p}{\rho} = e + \frac{u_i^2}{2} + \frac{p}{\rho} = h + \frac{u_i^2}{2}.$$

For perfect gas, the static enthalpy $h$ is assumed to be proportional to the temperature $T$:

$$h = c_p T,$$

where $c_p$ – specific heat capacity at constant pressure.

The term $Q_j$ expresses energy flux transferred by heat conductivity along the coordinate $x_j$ and is related to the temperature $T$ according to the Fourier's Law:

$$Q_j = -\lambda \frac{\partial T}{\partial x_j}, \tag{11}$$

where $\lambda$ – heat conductivity factor.

For flows in rotating frame of reference, instead of full enthalpy $H$, full rothalpy $I = H - 0.5\omega^2 r^2$ should be used in Eq. (10), where $\omega$ is rotational speed, $r$ is radius-vector magnitude.

When considering flow or incompressible or weakly compressible fluid, having introduced the energy equation, we introduce in the system of equations a new independent unknown parameter, the temperature. Density of liquid (and gas – at low flow velocities) does not depend on pressure and depends only on temperature.

The energy equation (10) allows for, e.g., simulation of heating of cold fluid flowing in a cavity with hot walls. Statement of problem requires specification of fluid temperature at the initial moment, including temperature at the inlet and outlet sections. As boundary conditions, temperature of cavity walls (if it is fixed during the considered process) or heat energy flux $Q_j$ through the walls (if heat energy flux is fixed) is specified.

The Eq. (10), written for time averaged values in order to simulate turbulent flows, looks as follows:

$$\frac{\partial}{\partial t}(\rho \overline{H}) + \frac{\partial}{\partial x_j}(\rho \overline{u_j H}) + \frac{\partial}{\partial x_j}(\rho \overline{u'_j H'}) = -\frac{\partial \overline{Q_j}}{\partial x_j} + \overline{f_i u_i}. \qquad (12)$$

Such form of presentation for equations (2), (3) and (12) suits for simulation of incompressible fluid flows. For compressible fluid flows, density is also subjected to fluctuations, and with the Reynolds time-averaging procedure, new unknown terms are obtained. In order to avoid this, the Favre-averaging procedure is used, i.e., instantaneous values of variables are not only time-averaged, but also mass-averaged. This procedure and the obtained equations are described, e.g., in [4].

**3. Simulation of natural convection flows.** Even in a vessel with initially stationary fluid, the flow may occur due to uneven heating resulted in uneven density distribution. The examples are flows in oceans and atmosphere. For their analysis, gravity force should be taken into account in Eq. (1). Natural convection flows are widely encountered also in engineering, in particular, in chemical industry. Natural convection should be also accounted for in designing of heat and ventilation systems for apartments.

**4. Simulation of compressible fluid flows.** In gas medium, temperature change can be caused not only by forced heating or cooling, but also, not to a lesser degree, by uneven distribution of density at high velocities. Density change can be caused not only by large temperature differences due to forced heating, but also by large pressure differences due to high flow velocities. Density of gas becomes variable though space at velocity about 0.3 Mach number and above. For simulation of so high-speed flows, the energy equation should be written in the following form:

$$\frac{\partial}{\partial t}(\rho H) + \frac{\partial}{\partial x_j}(\rho u_j H) = -\frac{\partial p}{\partial t} + \frac{\partial}{\partial x_j}\left[\mu u_i\left(\frac{\partial u_i}{\partial x_j} + \frac{\partial u_j}{\partial x_i}\right) - Q_j\right] + f_i u_i. \qquad (13)$$

For low-speed flows, this equation is reduced to (10). As is shown in the paper [4], order-of-magnitude analysis for this case allows for discarding terms with viscosity and pressure. However, when simulating high-speed flows (above 0.3 Mach number), these terms play substantial role.

When considering flow of compressible medium (i.e., medium, which density depends essentially on pressure differences), having introduced the energy equation (13), we introduce two new unknown independent parameters – temperature and density. For closing the resulting system of equations, the equation of state (8) is used. This equation specifies relation between these parameters.

In compressible fluid flows, pressure is contained in the system of equations not only in the form of first derivatives, but also explicitly – in the equation of state (8). Therefore, statement of problem requires specification of distribution of pressure (or density) at the initial moment.

The energy equation (13) allows for simulation of temperature variation caused not only by forced heating, but also by large pressure differences occurring at high velocities. For example, in wind tunnels, creation of supersonic flow is accompanied by substantial drop of temperature of medium, and this effect can be simulated using this equation.

As may be demonstrated, for compressible fluid flows, numerical behavior of system of model equations, with the Eq. (13) included, depends essentially on the Mach number $M$. Mach number is defined as ratio of average velocity of flow at some section to the acoustic speed *a* for this medium. The acoustic speed (sound speed) is defined as follows:

$$a = \sqrt{\frac{\partial p}{\partial \rho}}.$$

Acoustic speed is the speed of distribution of acoustic waves (compression waves of small amplitude) though the compressible medium. For ideal gas, this expression is reduced to the form

$$a = \sqrt{\gamma \frac{p}{\rho}},$$

where $\gamma = c_p / c_v$, i.e., the ratio of specific heat capacities at constant pressure and constant volume.

If flow velocities exceed the acoustic speed ($M > 1$), the system of model equations demonstrates hyperbolic behavior, thus not requiring to specify boundary conditions at the exit of computational domain. The obtained solutions are featured with so-called shock waves – surfaces of jump of flow parameters. The same phenomenon is observed in experiments. Along the streamlines crossing the shock wave, the conservation laws for mass, momentum and energy are fulfilled. These laws are known as conditions of Rankine – Hugoniot and are formulated, e.g., in [5, 6].

*5. Specifying dependencies of properties of medium on temperature.* When simulating fluid flows with heat transfer, as source data, of course, laws of variation of physical properties of the medium (density, viscosity, specific heat conductivity, heat capacity) on the temperature should be specified. These properties may be specified in form of arrays. Thus, for computing properties at certain temperature, e.g., piecewise-linear approximation can be used. Then, for example, the density will be computed using the following formula:

$$\rho(T) = \rho_n + \frac{\rho_{n+1} - \rho_n}{T_{n+1} - T_n}(T - T_n),$$

where $T$ – temperature in a certain point of space, $\rho(T)$ – corresponding density value, $T_n$, $T_{n+1}$ and $\rho_n$, $\rho_{n+1}$ – the nearest values of temperature, for which a user has specified the density, and corresponding values of density.

Viscosity, specific heat conductivity and specific heat capacity are specified and computed the same way. Density of gases depends not only on temperature, but also on pressure, and can be computed, using the equation of state (8). In modern CFD software tools, the database of most widely used substances (water, air, etc.) is already available, where dependencies of their physical properties on temperature and pressure are stored.

*6. Simulation of heat transfer in solid medium. Conjugate heat transfer.* The energy equation (10) can be used for simulation of heat transfer not only in fluids, but also in solid medium. As flow in this case is not available (i.e., $u = 0$), and full enthalpy is reduced to the form $H = c\,T$, where $c$ is specific heat capacity of solid material, the energy equation (10) appears as follows:

$$\frac{\partial}{\partial t}(\rho c T) = \frac{\partial}{\partial x_j}\left(\lambda \frac{\partial T}{\partial x_j}\right) + S_T, \qquad (14)$$

where $S_T$ – energy source inside solid body (if available).

Equation (14) is the sole equation required for simulation of heat transfer in solid medium. The only unknown variable in this equation is the temperature (i.e., distribution of temperature inside solid body). As boundary conditions required for solution of this equation, distribution of temperature on the surface of computational volume that corresponds to the solid body should be specified. Instead of temperature, heat flux $Q_j$ defined by Eq. (11) may be specified at a part of the surface.

In engineering practice, necessity to solve problems of conjugate heat transfer is often encountered. Such problems imply conjugate computation of fluid flows and computation of heat transfer in a solid thick-walled shell encompassing the fluid. In such cases, fluid flows are simulated, e.g., by equations (1) – (2) and (10), and heat transfer in solid body is modeled by Eq. (14). At the interface between fluid and solid phase, the condition of equality of temperatures and heat fluxes from the fluid and solid sides are stated. An example of heat transfer problem is computation of temperature field at the blades of a gas turbine, when those blades are forcedly cooled inside with cold fluid. Similar problem was solved, e.g., in [7].

## SIMULATION OF MULTI-PHASE FLOWS

*1. Introduction.* In engineering and environment, multi-phase flows of very different nature are met. They were reviewed, e.g., in [8]. We will distinguish the following types of multi-phase flows. First case – the considered volume is completely filled with a substance of one phase (e.g., liquid), and a substance of another phase in this volume is available only as discrete particles (of solid phase) or bubbles (of gaseous phase) which volume fraction is small (below 10% of total volume). Second case – the considered volume is partly filled with liquid and partly with gas, and the substances of these two phases are immiscible and separated one from another with free surface. Third case, the most complex – the substances of different phases can intermix (dissolve / deposit), and the volume fraction of the substance of second phase is large (above 10% of total volume). In order to simulate multi-phase flows of these types, different approaches are used, which are described below.

*2. Discrete phase model.* This approach is used for simulation of two-phase flows, when one of phases is presented in form of discrete particles, and volume fraction occupied by these particles is small (below 10% of total volume). Examples of such flows are water drops dispersed in airflow, air bubbles in liquid flow, solid particles in flow of air or water. The substance of main phase is supposed to be continuous medium, and its flow is simulated by Navier – Stokes (1) (or Reynolds (3)) and continuity equation. The substance presented in form of discrete particles does not form continuous medium; separate particles interact with flow of main phase and with each other discretely. In order to simulate motion of particles of the dispersed phase, Lagrangian approach is used, i.e., motion of each particle of the dispersed phase, under action of forces induced by flow of the main phase, is modeled individually.

This model is appropriate for simulation of flows, for example, in spray dryers, cyclone separators, fabric filters, at erosion of valves, at fuel injection in engines (www.adapco-online.com/adapco_online/uconf/nauc2002/mphase/index.html). The problems of such type often occur also in chemical industry, see, e.g., [9].

The particles of dispersed phase are assumed for simplicity to be of spherical shape. Forces acting on a particle are caused by the difference in velocity between the particle and the fluid of main phase, and by displacement of the fluid by the particle. The equation of motion for such a particle was deduced in [10] and looks as follows:

$$m_p \frac{dv_p}{dt} = 3\pi\mu d C_{cor}(v_f - v_p) + \frac{\pi d^3 \rho_f}{6}\frac{dv_f}{dt} + \frac{\pi d^3 \rho_f}{12}\left(\frac{dv_f}{dt} - \frac{dv_p}{dt}\right) +$$

$$+ F_e - \frac{\pi d^3}{6}(\rho_p - \rho_f)\vec{\omega}\times(\vec{\omega}\times\vec{r}) - \frac{\pi d^3 \rho_p}{3}(\vec{\omega}\times v_p). \qquad (15)$$

Here $m_p$ – particle mass, $d$ – particle diameter, $v$ – velocity, $\mu$ – dynamic viscosity of the fluid of main phase, $C_{cor}$ – its coefficient of viscous drag; $\vec{\omega}$ – rotational speed, $\vec{r}$ – radius-vector (when considering flow in a relative frame of reference). Index $p$ refers to a particle, index $f$ refers to the fluid of main phase.

Left-hand side of Eq. (15) represents sum of all the forces acting on the particle expressed through mass and acceleration of this particle. The first term at the right-hand side expresses deceleration of the particle due to viscous drag against the fluid flow, according to Stokes law. The second term is the force acting on the particle due to the pressure difference in the fluid around the particle caused by fluid acceleration. The third term is the force required for acceleration of virtual mass of fluid in the volume displaced by the particle. These two terms should be taken into account when density of fluid exceeds density of particles, e.g., when considering air bubbles in water flow. The forth term ($F_e$) is the external mass force that acts on the particle directly, e.g., gravity force or force of electric field. The last two terms are centrifugal force and Coriolis force, which are available only when considering motion in a relative frame of reference. Besides, sometimes at the right-hand side of (15) some additional forces should be taken into account (e.g., in flows with large temperature difference).

The Eq. (15) is first-order differential equation where the sole unknown value is the particle velocity $v_p$, and the argument is time $t$. Fluid flow velocity $v_f$ in all cells of the space is assumed to be known. As source data, except for size and properties of the particle, its initial position is specified. It is indicated also, what will occur with the particle when it hits the wall or another particle. For performing of computation, the terms containing $v_p$ are transferred to the left-hand side of Eq. (15). Particle velocity and position in each subsequent moment is determined by numerical integration of all the other terms in Eq. (15), with some time step $\Delta t$.

Coefficient of viscous drag $C_{cor}$ at moderate Reynolds number, $0.01 < Re_p < 260$, can be computed, e.g., as follows

$$C_{cor} = \begin{cases} 1 + 0.1315(\text{Re}_p)^{0.82-0.05\alpha} & at \ \text{Re}_p < 20 \\ 1 + 0.1935(\text{Re}_p)^{0.6305} & at \ \text{Re}_p > 20 \end{cases}$$

where $Re_p = \rho_f |v_f - v_p| d / \mu$, $\alpha = \log Re_p$.

In modern CFD software tools, the possibility is also available, to simulate heat and mass exchange between dispersed particle and fluid flow, e.g., evaporation of a drop at low enough fluid pressure or high enough temperature. The algorithms implemented in Fluent, CFX, STAR-CD, allows for simulation of influence on the fluid flow made by dispersed particles moving in this flow. As a first approximation, fluid density and viscosity, as well as some other parameters, are multiplied by $(1 - \alpha_p)$, where $\alpha_p$ is specific volume occupied by discrete particles. Then, at each further time step, changes in mass, momentum and energy of discrete particles are computed, and the obtained differences are added to the mass (2), momentum (1) and energy (10) conservation equations for fluid flow of the main phase, in form of source terms. Thus, computation of fluid flow of the main phase and computation of motion of dispersed particles is performed jointly.

If fluid flow of the main phase is turbulent, then trajectory of dispersed particles is not predetermined and depends on intensity and direction of turbulent fluctuations. An approach for simulation of impact of turbulent fluctuations of fluid flow on motion of dispersed particles was suggested, in particular, in [11].

In modern CFD software tools, several boundary conditions are implemented that correspond to different events occurring when a discrete particle hits solid wall. The possible events are as follows: recoil due to elastic or inelastic impact, adhesion to the wall, creeping along the wall (in dependence on physical properties and angle of incidence), passing through the wall (if the wall is porous), etc. A possibility is also available, to simulate splitting and merging of water drops or air bubbles when they hit one another, under certain conditions.

***3. Model of free surface flow.*** This approach allows for simulation of flow of two (or more) liquids or combination of liquid and gas that do not intermix and, being exposed to mass forces, form the clearly expressed interface, i.e., free surface.

According to this approach, in order to simulate free surface, the system of model equations is supplemented with the equation of transfer of the fill function $F$ that expresses "concentration of liquid in gas" (when considering gas-liquid flow). This implies the name of the model of flow – VOF model (Volume Of Fluid, i.e., volume fraction occupied by fluid).

$$\frac{\partial (F\rho)}{\partial t} + \frac{\partial}{\partial x_j}(F\rho u_j) = 0 . \tag{16}$$

$F$ equals 1 in the volume occupied by fluid and 0 in the volume occupied by gas. Only for cells crossed by the free surface, $0 < F < 1$. Initial position of the free surface should be specified. The algorithm for numerical solution using this model is described in [12, 13].

Examples of problems where such approach is appropriate: filling of fuel tanker, splashing or boiling in a vessel with free surface, simulation of flow around sea ships. In the paper [14], in particular, the CFD software tool FlowVision with VOF model was used for simulation of water flow around the car wheel driving in a puddle in the aquaplaning mode.

**4. Multiphase mixture model.** This approach allows for simulation of multi-phase flows when the substances of different phases can intermix and do not form the free surface.

In order to simulate the flow of two or several phases, this model uses one continuity equation, one set of momentum equations and one energy equation that are written with regard to mass-averaged velocity and density of the mixture. Thus, the continuity equation in this model looks as follows:

$$\frac{\partial \rho_m}{\partial t} + \frac{\partial}{\partial x_j}\left(\rho_m u_{mj}\right) = m, \qquad (17)$$

where $\rho_m$ is density of the mixture (Eq. (5)), $u_m$ is mass-averaged velocity, $u_{mj}$ is velocity projection to the axis $x_j$. On the index $j$, the summation is assumed. The term $m$ represents mass transfer due to cavitation and/or other physical effects. By default, it equals zero.

This model uses the conception of drift velocities to take into account that motion of different phases occurs with different velocities. This allows for simulation, e.g., deceleration of sand grains flying into a vessel filled with still fluid.

In this model, the momentum equation projected to the axis $x_i$ looks as follows:

$$\frac{\partial}{\partial t}\left(\rho_m u_{mi}\right) + \frac{\partial}{\partial x_j}\left(\rho_m u_{mi} u_{mj}\right) = -\frac{\partial p}{\partial x_i} + \frac{\partial}{\partial x_j}\left[\mu_m\left(\frac{\partial u_{mi}}{\partial x_j} + \frac{\partial u_{mj}}{\partial x_i}\right)\right] + f_i + \frac{\partial}{\partial x_i}\left(u_{ki} - u_{mi}\right)^2, (18)$$

where $\mu_m$ is the mass-averaged viscosity (Eq. (6)), $u_k$ is the velocity of substance of $k$-phase, $u_{mj}$ is the projection of this velocity to the axis $x_i$, $(u_k - u_m)$ is the slip velocity of substance of $k$-phase relative to the mass-averaged velocity $u_m$. Eq. (18) defers from Eq. (1) by presence of the last term that simulates mutual slip of velocities.

The energy equation is written the same way.

The continuity equation for a separate secondary phase $k$ can be presented as follows

$$\frac{\partial (F_k \rho_k)}{\partial t} + \frac{\partial}{\partial x_j}\left(F_k \rho_k u_{mj} + F_k \rho_k u_{kj}\right) = 0, \qquad (19)$$

where $\rho_k$ is the density of substance of $k$-phase. From this equation, volume fraction $F_k$ occupied by the substance of $k$-phase in a certain cell of the space can be determined.

**5. Multi-phase Eulerian model.** This model is the most general and the most complex among all the models of multi-phase flow. The substance of each phase is assumed to form continuous medium. Its motion is simulated with own system of Navier – Stokes (Reynolds) equations, continuity equation and energy equation.

According to this model, the equations written for each phase are solved jointly. An algorithm for computation of such flows was suggested, in particular, in [15], and implemented in CFX, Fluent and STAR-CD (see, e.g., www.adapco-online.com). This model requires the most of computer resources, both concerning RAM volume and processor speed.

This model is appropriate for simulation of flows, e.g., in 2-phase mixing vessel (gas-liquid mixer), fluidized bed, settling tank, gas-lift reactor, liquid-liquid extraction column (see www.adapco-online.com/adapco_online/uconf/nauc2002/mphase/index.html).

**6. Additional recommendations for selection of a multi-phase flow model.** In order to simulate flows where the substances of different phases intermix and do not form the interface, in many cases both discrete phase model, mixture model and Eulerian model can be used. Additional criteria for selection of the most proper model are as follows [16].

- *Mass density ratio β of the dispersed phase (d) to that of the carrier phase (c)*:

$$\beta = \gamma \frac{F_d}{F_c}, \qquad (20)$$

where $F_d$ and $F_c$ are volume fractions, $\gamma$ is ratio of densities of the dispersed and carrier phase, $\gamma = \rho_d / \rho_c$; this ratio can be above 1000 for solid particles in gas flow, about 1 for solid particles in liquid flow, and below 0.001 for gas particles in liquid flow.

When the ratio $\beta$ is very low, the dispersed particles almost do not influence the carrier flow; thus, any of the listed models can be used. At very high values of $\beta$, the dispersed particles strongly influence the carrier flow, and only the multi-phase Eulerian model should be used for adequate simulation of such flows. At moderate values of $\beta$, in order to select the proper model, it is necessary to calculate the Stokes number as follows.
- *Stokes number St*:

$$St = \frac{t_d}{t_c}, \tag{21}$$

where $t_d$ is characteristic time of particle motion, $t_d = (\rho_d\, d_d^2) / (18\, \mu_c)$, $d_d$ is particle diameter, $\mu_c$ is viscosity of substance of the carrier phase, $t_c = L_c / U_c$ is characteristic time of carrier flow, $L_c$ is characteristic length, $U_c$ is characteristic velocity.

At $St \ll 1.0$, dispersed particles almost do not deflect from the carrier streamlines; thus, any of multi-phase models can be used (as a rule, the mixture model, as the least resource consuming). At $St > 1.0$, trajectories of dispersed particles do not coincide with the carrier streamlines at all. Thus, the mixture model is inapplicable here; either the discrete phase model or the Eulerian model should be used.

*7. Simulation of flows with cavitation.* When, at constant temperature, the pressure of liquid decreases below the saturated vapor pressure, the liquid loses continuity, and bubbles filled with vapor are formed in it. Besides, the liquid can contain other dissolved gases, in form of micro-bubbles. When the pressure decreases further, these bubbles can grow and form extensive cavities. If the pressure increases, these bubbles and cavities will collapse, vanishing and disappearing in the liquid. This entire process is called cavitation. Cavitation is accompanied with very strong and abrupt jumps of density, pressure and temperature; the flow becomes very unstable. The phenomenon of cavitation was described in detail, e.g., in [17].

In the modern CFD software tools, different approaches for simulation of cavitation are used. One of the most promising approaches was suggested, e.g., in [18]. In order to simulate cavitating flows, as a rule, the free surface model or the mixture model is used. The Eulerian model is hardly expedient to use, due to low values of parameters $\beta$ and $St$ (Eq. (20) and (21)) in such flows.

It is assumed that the liquid, its vapor and other gases with specified physical properties may be present in the computational domain. In the model equations of the mixture model, some terms that represent heat and mass transfer due to cavitation are introduced. These terms were suggested, in particular, in [18].

## OTHER POSSIBILITIES OF SIMULATION OF FLUID FLOWS

In the modern CFD software tools, a number of other possibilities for simulation of fluid flows are implemented. Some of them are briefly described below.

*1. Non-Newtonian fluid flows.* In common (Newtonian) fluids, shear stress $\tau$ in a certain liquid volume obeys the Newtonian law, i.e., is proportional to the rate-of-deformation tensor $D$ for that volume, $\tau = \mu\, D$, where

$$D = \left( \frac{\partial u_j}{\partial x_i} + \frac{\partial u_i}{\partial x_j} \right),$$

$\mu$ – dynamic viscosity coefficient that do not depend on $D$.

In non-Newtonian fluids, to which belong, e.g., paraffin, wax, honey, tar, the viscosity coefficient $\mu$ depends on $D$.

The dependence $\mu\,(D)$ should be specified as source data for computation of flows. By now, a number of approximate laws were suggested that describe this dependence for different non-Newtonian fluids. This topic is described with more details, e.g., in [19].

*2. Flows through porous medium. Flows with distributed resistance.* In some cases, fluid flows are hindered by a host of small solid obstacles that influence the flows substantially but are too small in size to be modeled individually. Examples of such flows in the nature are air motion though a forest, water flow in a river stuffed with algae. In engineering: liquid or gas flows though a porous filter element, gland, perforated plate or honeycomb.

For simulation of such flows, as a rule, a host of small solid elements that obstacle the flow are replaced by the entire uniformly distributed resistance. In order to simulate this resistance, an additional source term $S_i$ is introduced to the right-hand side of the momentum equations (1):

$$S_i = -\left(\sum_{j=1}^{3} D_{ij}\mu u_{ij} + \sum_{j=1}^{3} C_{ij}\frac{1}{2}\rho|\vec{u}|u_{ij}\right), \tag{22}$$

where $D_{ij}$ and $C_{ij}$ are matrixes 3 x 3 predefined at problem statement.

In case of uniform (isotropic) porous medium, diagonal elements of the matrix $D_{ij}$ equal $1/\alpha$, diagonal elements of the matrix $C_{ij}$ equal $C_2$, and all the other elements of these matrixes equal zero. Here $\alpha$ is permeability; $C_2$ is internal resistance coefficient. Then $S_i$ looks as follows:

$$S_i = -\left(\frac{\mu}{\alpha}u_i + C_2\frac{1}{2}\rho|\vec{u}|u_i\right). \tag{23}$$

At low velocities of flow, when the flow is laminar, only the first part of the term $S_i$ is essential. When neglecting the convection acceleration and diffusion, the equation of motion though the porous medium is reduced to Darcy's Law:

$$\frac{\partial p}{\partial x_i} = -\frac{\mu}{\alpha}u_i.$$

At high velocities of flow, on the contrary, only the second part of the term $S_i$ is essential. The coefficient $C_2$ may be understood as loss factor per unit of length in the flow direction. Thus, pressure drop may be presented as the function of dynamic head:

$$\frac{\partial p}{\partial x_i} = -C_2\frac{1}{2}\rho|\vec{u}|u_i.$$

Heat conductivity of the porous medium $\lambda_{eff}$ can be computed as the volume-averaged heat conductivity of substances of fluid $\lambda_f$ and solid $\lambda_s$ phases in the porous medium:

$$\lambda_{eff} = \gamma\lambda_f + (1-\gamma)\lambda_s,$$

where $\gamma$ is porosity of the medium, i.e., volume fraction occupied by the fluid medium.

***3. Flows with radiating heat transfer.*** As it is known, there are three ways of heat transfer: by convection, by diffusion and by radiation. The energy equation (10) or (13) allows for simulation of only two former types of heat transfer. In many engineering problems, radiating heat transfer may be neglected. However, it can play the dominating role when very high temperature differences are present. Radiating heat flux $Q_{rad}$, e.g., between hot and cold walls, is expressed as $Q_{rad} = \sigma(T_{max}^4 - T_{min}^4)$, where $T_{max}$ and $T_{min}$ – temperature of hot and cold walls, $\sigma = 5.67 \cdot 10^8$ W / (m$^2$ K$^4$) – Stefan-Boltzmann constant. From the physical point of view, radiating heat transfer is the flow of photons of certain frequency range. Like light beams, radiating heat is transferred along straight lines.

Fluid flowing in the volume subjected to radiating heat transfer is heated from hot walls by convection and diffusion. Besides, if the fluid medium is not completely transparent, radiating heat is transferred to the flow also directly. In order to simulate this process, a new term is added to the energy equation (10) or (13). Appearance of this term is determined according to the assumed model of radiating heat transfer. Review of modern models of radiating heat transfer is presented, e.g., in [20].

An example of problem that demonstrates the radiating heat transfer is included in the demo-version of CFD software tool FlowVision, which can be freely downloaded at www.flowvision.ru.

***4. Flow with chemical reactions and combustion.*** In the system of model equations, the equations of concentration transfer (4) for each component of the mixture, and the energy equation (10) or (13) should be included. Besides, the equations for simulation of chemical reaction or combustion process are introduced. For a number of typical reactions (in particular, methane-air burning), the corresponding models are already developed and implemented in modern CFD software tools. Besides, these software tools contain some interface for user programming. Thus, experienced enough users can create an appropriate model for some specific reaction.

Combustion flows usually feature with very strong turbulent fluctuations that influence essentially the flow pattern. In order to simulate this phenomenon, e.g., in FlowVision, $k - \varepsilon$ turbulence model is supplemented with a special equation for modeling of fluctuations.

*5. Prediction of acoustics.* In a number of modern CFD software tools, a possibility for prediction of intensity and frequency characteristics of aerodynamically generated noise is implemented. As source data for such prediction, the results of computation of fluid flow with non-stationary Navier – Stokes (or Reynolds) equations are required.

In order to predict acoustics properly, the flow should be computed using LES or DNS [2] turbulence model, because these models ensure the best resolution of pressure fluctuations occurring in the flow. The matter is, these fluctuations are the main source of aerodynamically generated noise.

## CONCLUSION

A review of fluid flow models implemented in the leading CFD software tools is presented. The described models are designed for simulation of multi-component and multi-phase flows, compressible flows, flows with heat transfer, cavitation and other phenomena. As is demonstrated in a number of publications, these software tools (CFX, Fluent, STAR-CD, etc.) allow for adequate simulation of complex physical effects of different nature, even for problems where performing of physical experiment is extremely difficult.